\documentclass{revtex4}
\usepackage{epsfig}
\usepackage{amssymb}
\usepackage{amsmath}
\usepackage{amsfonts}
\usepackage{graphicx}
\usepackage{mathrsfs}
\usepackage[dvips]{color}
\usepackage{multirow}

\newcommand{\R}{\mathbb{R}}

\newcommand{\fa}{\mathfrak{a}}

\newcommand{\fn}{\mathfrak{n}}

\newcommand{\fK}{\mathfrak{K}}

\newcommand{\bM}{\mathbf{M}}

\newcommand{\bS}{\mathbf{S}}

\newcommand{\cP}{\mathcal{P}}

\newcommand{\cT}{\mathcal{T}}

\newcommand{\be}{\begin{equation}}
\newcommand{\ee}{\end{equation}}
\newcommand{\bea}{\begin{eqnarray}}
\newcommand{\eea}{\end{eqnarray}}
\newcommand{\nn}{\nonumber}

\newcommand{\ed}{\end{document}}

\newcommand{\rz}{{\rm z}}

\newcommand{\bi}{\begin{itemize}}
\newcommand{\ei}{\end{itemize}}

\newcommand{\bce}{\begin{center}}
\newcommand{\ece}{\end{center}}

\newcommand{\RE}{\,{\rm Re}}
\newcommand{\IM}{\,{\rm Im}}

\begin{document}
\title{Invisibility and $\cP\cT$-symmetry}
\author{Ali~Mostafazadeh}
\address{Department of Mathematics, Ko\c{c}
University, Sar{\i}yer 34450, Istanbul, Turkey\\
amostafazadeh@ku.edu.tr}

\begin{abstract}

For a general complex scattering potential defined on a real line, we show that the equations governing invisibility of the potential are invariant under the combined action of parity and time-reversal ($\cP\cT$) transformation. We determine the $\cP\cT$-symmetric as well as non-$\cP\cT$-symmetric invisible configurations of an easily realizable exactly solvable model that consists of a two-layer planar slab consisting of optically active material. Our analysis shows that although $\cP\cT$-symmetry is neither necessary nor sufficient for the invisibility of a scattering potential, it plays an important role in the characterization of the invisible configurations. A byproduct of our investigation is the discovery of certain configurations of our model that are effectively reflectionless in a spectral range as wide as several hundred nanometers.
\medskip

\hspace{6.2cm}{Pacs numbers: 03.65.-w, 03.65.Nk, 42.25.Bs,
24.30.Gd}
\end{abstract}

\maketitle

\section{Introduction}

Consider a scattering problem for a real or complex potential $v$ that is defined on the real line. The asymptotic solution of the time-independent Schr\"odinger equation has the form
    \be
    \psi(z)\to A_\pm e^{ikz}+B_\pm e^{-ikz}~~{\rm as}~~z\to \pm\infty,
    \label{psi}
    \ee
where $k$ is the wavenumber that takes real and positive values, and $A_\pm$ and $B_\pm$ are possibly $k$-dependent complex coefficients. We can obtain the information about the scattering properties of $v$ using its transfer matrix $\bM$ that is defined by the equation:
    \be
    \left[\begin{array}{c}A_+\\ B_+\end{array}\right]=\bM
    \left[\begin{array}{c}A_-\\ B_-\end{array}\right].
    \label{M-def}
    \ee
It is not difficult to show that $\bM$ has a unit determinant \cite{jpa-2009}. This in turn implies that for every real or complex scattering potential the left and right transmission amplitudes coincide while this is not generally true for the left and right reflection amplitudes \cite{prl-2009}. Let $T$ and $R^{r,l}$ respectively denote the (complex) transmission and right/left reflection amplitudes, so that the transmission and right/left reflection coefficients are given by $|T|^2$ and $|R^{r,l}|^2$. We can express $\bM$ in terms of $T$ and $R^{r,l}$ according to \cite{prl-2009}
    \be
    \bM=\left[\begin{array}{cc}
    \displaystyle T-\frac{R^lR^r}{T} & \displaystyle \frac{R^r}{T} \vspace{.2cm}\\ 
    \displaystyle -\frac{R^l}{T} & \displaystyle \frac{1}{T}\end{array}\right].
    \label{M}
    \ee
Alternatively, we can describe the scattering properties of the system using its $S$-matrix,
    \be
    \bS:=\left[\begin{array}{cc}
    T & R^r\\
    R^l & T\end{array}\right].
    \label{S}
    \ee

The potential is called \emph{reflectionless from the left} (resp.\  \emph{right)}, if $R^l=0$ and $R^r\neq 0$ (resp.\ $R^r=0$ and $R^l\neq 0$.) As seen from (\ref{M}) and (\ref{S}), unidirectional reflectionlessness implies the non-diagonalizability of both $\bM$ and $\bS$. Therefore, the parameters of the potential for which it becomes unidirectionally reflectionless correspond to exceptional points  \cite{ep} of $\bM$ and $\bS$.

The potential is called \emph{invisible from the left} (resp.\ \emph{right}), if it is reflectionless from left (resp.\ right) and in addition $T=1$. Therefore, in light of (\ref{M}), we can express the condition for the invisibility of the potential from the left in terms of the entries of the transfer matrix according to
    \be
    M_{12}\neq 0,~~~~~M_{21}=0,~~~~~M_{11}=M_{22}=1.
    \label{invisible}
    \ee
Similarly, for invisibility from the right we have
    \be
    M_{12}=0,~~~~~M_{21}\neq 0,~~~~~M_{11}=M_{22}=1.
    \label{invisible-right}
    \ee
Because $\det(\bM)=1$, Equations~(\ref{invisible}) are not independent. The same holds for (\ref{invisible-right}). We can reduce (\ref{invisible}) and (\ref{invisible-right}) to the following sets of equivalent independent equations respectively.
    \bea
    &&M_{12}\neq 0,~~~~~M_{11}=M_{22}=1.
    \label{invisible3}\\
    &&M_{21}\neq 0,~~~~~M_{11}=M_{22}=1.
    \label{invisible4}
    \eea

Recently Lin~ et al \cite{unidir} have reported the emergence of unidirectional invisibility for complex $\cP\cT$-symmetric locally periodic potentials of the form
    \be
    v(z):=\left\{\begin{array}{ccc}
    \alpha_0+\alpha\, e^{2i\beta z} &{\rm for} & |z|\leq\frac{L}{2},\\[6pt]
    0  &{\rm for} & |z|>\frac{L}{2},\end{array}\right.
    \label{lin=}
    \ee
where $\alpha_0,\alpha,\beta,L$ are real parameters, and $\beta$ and $L$ are positive. The spectral properties of the periodic potentials: $v(z)=\sum_{n=0}^\infty \alpha_n e^{i\beta_n z}$, have been studied by Gasymov more than three decades ago \cite{gasymov}. The special case $v(z)=\alpha e^{i\beta z}$ has been reexamined more recently in \cite{CM-jmp-2007,midya,graefe}. See also \cite{uzdin}. The publication of \cite{unidir} has motivated a more detailed study of  the potential (\ref{lin=}). In particular, the authors of  \cite{longhi-jpa, jones} use the analytic solution of the the wave equation \cite{gasymov,CM-jmp-2007} to improve the approximate results of \cite{unidir} that rely on the rotating-wave approximation. For an earlier discussion of reflectionlessness and $\cP\cT$-symmetry, see \cite{abb}.

In the present paper, we provide a thorough assessment of the role of $\cP\cT$-symmetry \cite{PT-sym} in the phenomenon of invisibility of scattering potentials in one dimension, and explore the invisible configurations of an exactly solvable model that admits experimental realizations.

\section{$\cP\cT$-Symmetric Nature of Invisibility}

We begin our analysis by examining the combined effect of parity and time reversal ($\cP\cT$) transformation on the transfer matrix $\bM$. It is easy to show that under this transformation the asymptotic solutions of the wave equation (\ref{psi}) transform according to
    \be
    \psi(z)\stackrel{\cP\cT}{\longrightarrow}(\cP\cT\psi)(z):=\psi(-z)^*\to A_\pm^*e^{ikz}+B_\pm^*e^{-ikz}~~~
    {\rm for}~~~ z\to\mp\infty,
    \label{psi-PT}
    \ee
where we suppose that $k$ is real and positive. The transformation rule (\ref{psi-PT}) implies that the transfer matrix of the $\cP\cT$-transformed system, that we denote by $\bM^{(\cP\cT)}$, satisfies
    \be
    \left[\begin{array}{c}A_-^*\\ B_-^*\end{array}\right]=\bM^{(\cP\cT)}
    \left[\begin{array}{c}A_+^*\\ B_+^*\end{array}\right].
    \label{M-PT}
    \ee
Combining (\ref{M-def}) and (\ref{M-PT}), we find the following transformation rule for the transfer matrix.
    \be
    \bM\stackrel{\cP\cT}{\longrightarrow}\bM^{(\cP\cT)}:=\bM^{-1*}.
    \ee
Because $\det\bM=1$, this means that
    \be
    M_{11}\stackrel{\cP\cT}{\longrightarrow} M_{22}^*,~~~~~
    M_{12}\stackrel{\cP\cT}{\longrightarrow} -M_{12}^*,~~~~~
    M_{21}\stackrel{\cP\cT}{\longrightarrow} -M_{21}^*,~~~~~
    M_{22}\stackrel{\cP\cT}{\longrightarrow} M_{11}.
    \label{M-PT-trans}
    \ee
In particular, as noted in \cite{longhi1}, the transfer matrix of a $\cP\cT$-symmetric potential satisfies $\bM^{-1}=\bM^*$.

A straightforward consequence of (\ref{M-PT-trans}) is that in general, regardless of whether the potential is $\cP\cT$-symmetric or not, the equations governing its invisibility (both from left and right), namely (\ref{invisible3}) and (\ref{invisible4}), are invariant under the $\cP\cT$-transformation \footnote{Here by being invariant we mean that they are mapped to equations obtained by complex-conjugation or multiplication with a nonzero constant, so that they have the same solutions.}. We can restate this fact as follows.
    \begin{itemize}
    \item[]{\textbf{Invisibility Theorem:}} \emph{Consider a general real or complex scattering potential $v$ that is defined on $\R$. Let $v^{(\cP\cT)}$ be the $\cP\cT$-transform of $v$ that is given by $v^{(\cP\cT)}(z):=v(-z)^*$, and $k_\star$ be a positive real number. Then the following equivalent statements hold.}
            \begin{itemize}
            \item[ i)] \emph{$v$ is invisible from the left (or right) for $k=k_\star$ if and only if so is $v^{(\cP\cT)}$.}
            \item[ii)] \emph{$v$ is invisible from the left (resp.\ right) for $k=k_\star$ if and only if $v^*$ is invisible from the right (resp.\ left) for $k=k_\star$.}
            \end{itemize}
    \end{itemize}
For non-$\cP\cT$-symmetric potentials this theorem implies a pairing of the left/right-invisible configurations that are related by the $\cP\cT$-transformation. We also have a pairing of all complex (non-real) potentials that display unidirectional invisibility at a given wavenumber where each member of the pair is the time-reversal (complex-conjugate) of the other. Obviously, parity reflection transformation changes a left- (resp.\ right-) invisible potential to a right- (resp.\ left-) invisible potential. In particular, even ($\cP$-symmetric) potentials do not support unidirectional invisibility. In view of the above theorem, the same holds for real ($\cT$-symmetric) potentials. Unlike $\cP$-symmetric and $\cT$-symmetric potentials, $\cP\cT$-symmetric potentials can display unidirectional invisibility.

We can use the argument leading to the proof of the Invisibility Theorem to show that its statement also holds, if we replace the term ``invisible'' with ``reflectionless.'' In particular, we have $\cP\cT$-dual and $\cT$-dual pairings of complex reflectionless potentials.

\section{Two-Layer Infinite Slab Model}

\begin{figure}
    \begin{center}
    \includegraphics[scale=.6,clip]{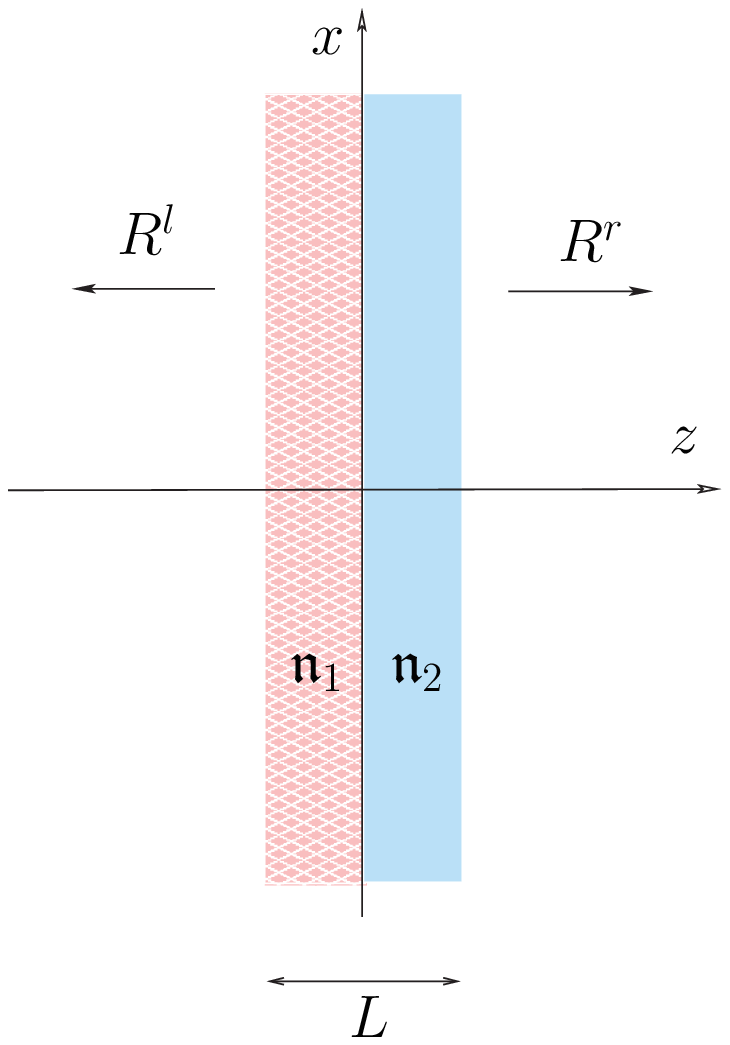}
    \caption{(Color online) Schematic view of the cross-section of a two-layer infinite planar slab of gain material of thickness $L$ that is aligned in the $x$-$y$ plane. The arrows marked by $R^l$ and $R^r$ represent the reflected waves associated with incident waves from the left and right, respectively. \label{fig0}}
    \end{center}
    \end{figure}

Consider an infinite planar two-layer slab of optically active material as depicted in Figure~\ref{fig0}.
Suppose that the layers have equal thickness, $L/2$, and different gain/loss properties. Then the wave equation for a linearly polarized time-harmonic electromagnetic wave that  propagates along the normal direction to the slab can be reduced to the Helmholtz equation \cite{p108},
    \be
    \Psi''(z)+k^2\fn(z)^2\Psi(z)=0,
    \label{HH}
    \ee
where
    \be
    \fn(\rz):=\left\{\begin{array}{ccc}
    \fn_1 & {\rm for} & -\frac{L}{2}\leq \rz <0,\\[6pt]
    \fn_2 & {\rm for} & 0\leq \rz \leq \frac{L}{2},\\[6pt]
    1 & {\rm for} & |\rz|>\frac{L}{2},
    \end{array}\right.
    \label{n=}
    \ee
and $\fn_1$ and $\fn_2$ are complex refractive indices of the layers. In Ref.~\cite{p108}, we have used this model to examine the importance of $\cP\cT$-symmetry for generating self-dual spectral singularities, i.e., spectral singularities \cite{prl-2009} that coincide with their time-reversal dual. The concept of a self-dual spectral singularity provides the mathematical basis for an optical device that functions as a coherent perfect absorber (CPA) \cite{antilaser,longhi1} whenever it is subject to incidence coherent waves with identical amplitude and phase from both sides and operators as a laser otherwise. It is therefore called a CPA-laser \cite{stone}. In \cite{p108}, we establish the existence of non-$\cP\cT$-symmetric CPA-lasers. Here we use the optical system defined by (\ref{n=}) to examine the role of $\cP\cT$-symmetry in achieving unidirectional invisibility.

We can calculate the transfer matrix $\bM$ of the above model using the method outlined in \cite{pra-2011b} and employed in \cite{p108}. Doing the necessary calculations this yields the following expressions for the entries of $\bM$.
     \bea
    M_{11}&=& (\fn_+^2-\fn_-^2)^{-1}
    \Big\{\fn_+^2\cos\fa_+-\fn_-^2\cos\fa_-+i(\tilde\fn_+\fn_+\sin\fa_++\tilde\fn_-
    \fn_-\sin\fa_-)\Big\}\,e^{-i\fK},
    \label{M11}\\
    M_{12}&=&(\fn_+^2-\fn_-^2)^{-1} \Big\{\fn_-\fn_+(\cos\fa_+-\cos\fa_-)+
    i(\tilde\fn_+\fn_-\sin\fa_- +\tilde\fn_-\fn_+\sin\fa_+)\Big\},
    \label{M12}\\
     M_{21}&=&(\fn_+^2-\fn_-^2)^{-1} \Big\{\fn_-\fn_+(\cos\fa_+-\cos\fa_-)-
    i(\tilde\fn_+\fn_-\sin\fa_- +\tilde\fn_-\fn_+\sin\fa_+)\Big\},
     \label{M21}\\
     M_{22}&=& (\fn_+^2-\fn_-^2)^{-1}
    \Big\{\fn_+^2\cos\fa_+-\fn_-^2\cos\fa_--i(\tilde\fn_+\fn_+\sin\fa_++
    \tilde\fn_-\fn_-\sin\fa_-)\Big\}e^{i\fK},
    \label{M22}
    \eea
where
    \be
    \fn_\pm:=\fn_1\pm\fn_2,~~~~~
    \fa_\pm:=\frac{\fn_\pm\fK}{2},~~~~~\fK:=Lk,~~~~~
    \tilde\fn_\pm:=\fn_1\fn_2\pm1=\frac{1}{4}(\fn_+^2-\fn_-^2)\pm 1.
    \label{mp}
    \ee
As a nontrivial check on the validity of equations~(\ref{M11}) -- (\ref{M22}), we have shown by direct calculations that indeed $\det\bM=1$.

In view of (\ref{M12}) -- (\ref{M22}), the condition (\ref{invisible}) for the invisibility of the system from the left takes the form:
    \bea
    &&\fn_-\fn_+(\cos\fa_+-\cos\fa_-)+
    i(\tilde\fn_+\fn_-\sin\fa_- +\tilde\fn_-\fn_+\sin\fa_+)\neq 0,
    \label{e1}\\
    &&\fn_-\fn_+(\cos\fa_+-\cos\fa_-)-
    i(\tilde\fn_+\fn_-\sin\fa_- +\tilde\fn_-\fn_+\sin\fa_+)=0,
    \label{e2}\\
    &&\fn_+^2\cos\fa_+-\fn_-^2\cos\fa_-+i(\tilde\fn_+\fn_+\sin\fa_++\tilde\fn_-
    \fn_-\sin\fa_-)=(\fn_+^2-\fn_-^2) e^{i\fK},
    \label{e4}\\
    &&\fn_+^2\cos\fa_+-\fn_-^2\cos\fa_--i(\tilde\fn_+\fn_+\sin\fa_++
    \tilde\fn_-\fn_-\sin\fa_-)=(\fn_+^2-\fn_-^2) e^{-i\fK}.
    \label{e3}
    \eea
Note that for $\fn_1=\fn_2$, $\fn_-=0$ and we can use Equations~(\ref{e2})  and (\ref{e3}) to show that in this case $\fn_1=1$. This corresponds to an empty slab. Therefore, without loss of generality, we take $\fn_2\neq\fn_1$. Moreover, by adding and subtracting Equations~(\ref{e3}) and (\ref{e4}), we can reduce them to
    \bea
    &&\fn_+^2\cos\fa_+-\fn_-^2\cos\fa_-=(\fn_+^2-\fn_-^2) \cos\fK,
    \label{e3a}\\
    &&\tilde\fn_+\fn_+\sin\fa_++\tilde\fn_-
    \fn_-\sin\fa_-=(\fn_+^2-\fn_-^2)\sin\fK.
    \label{e4a}
    \eea
Similarly using (\ref{e1}) and (\ref{e2}), we have
    \bea
    &&\fn_-\fn_+(\cos\fa_+-\cos\fa_-)\neq 0,
    \label{e11}\\
    &&\tilde\fn_+\fn_-\sin\fa_- +\tilde\fn_-\fn_+\sin\fa_+\neq 0.
    \label{e12}
    \eea
For non-exodic material, $\RE(\fn_{1,2})\geq 1$ and $\fn_+\neq 0$.
We also have $\fn_-\neq 0$. Therefore,  (\ref{e11}) holds if and only if
    \be
    \cos\fa_+\neq \cos\fa_-.
    \label{e13}
    \ee
We arrive at the same conclusion if we demand that the potential be invisible from the right. Therefore, a necessary and sufficient condition for the unidirectional invisibility of our system is that we satisfy (\ref{e3a}), (\ref{e4a}), and (\ref{e13}). To determine the direction of invisibility we need to check which of $M_{12}$ and $M_{21}$ vanishes.

\section{Bidirectional Invisibility}

Suppose that $\fn_1\neq\fn_2$ and the system is invisible from both directions, i.e., it satisfies (\ref{e3a}), (\ref{e4a}), and $\cos\fa_+=\cos\fa_-$. Substituting this relation in (\ref{e2}) and (\ref{e3a}) and noting that $\fn_+^2-\fn_-^2\neq 0$ gives
    \bea
    &&\cos\fa_\pm=\cos\fK,
    \label{ee1}\\
    &&\tilde\fn_+\fn_-\sin\fa_- +\tilde\fn_-\fn_+\sin\fa_+=0,
    \label{ee2}
    \eea
Equation~(\ref{ee1}) means that
    \be
    \fa_+=2\pi m_+\pm\fK,~~~~~~~~~~~~\fa_+=2\pi m_-\pm\fK,
    \label{ee10}
    \ee
where $m_\pm$ are integers. In view of (\ref{mp}), this implies that $\fn_\pm$, $\tilde\fn_\pm$, $\fn_1$, and $\fn_2$ must take real values. Another consequence of (\ref{ee1}) is
    \be
    \sin\fa_+=\pm\sin\fa_-.
    \label{ee3}
    \ee
Inserting this relation in (\ref{e2}) and (\ref{e3}) and using the fact that $\fn_\pm$, $\tilde\fn_\pm$, and $\fa_\pm$ are real, we find
    \[ (\tilde \fn_+\fn_-\pm\tilde\fn_-\fn+)\sin\fa_+=0,~~~~~
        (\tilde \fn_+\fn_+\pm\tilde\fn_-\fn-)\sin\fa_+=0.\]
Adding these two equations gives $2\fn_1(\tilde\fn_+\pm\tilde\fn_-)\sin\fa_+=0$. Because $\fn_1(\tilde\fn_+\pm\tilde\fn_-)\neq 0$ and (\ref{ee3}) holds, this implies that $\sin\fa_\pm=0$. Equivalently, there are integers $m'_\pm$ such that
    \be
    \fa_\pm=\pi m'_\pm,
    \label{ee4}
    \ee
Next, we combine this relation and (\ref{ee10}) to obtain
    \be
    \fK=\pi m,
    \label{ee5}
    \ee
where $m$ is an integer. In light of (\ref{ee4}) and (\ref{ee5}), we also find that $m_+'\pm m_-'$ are even, and
    \be
    \fn_1=\frac{2m_1}{m},~~~~~\fn_2=\frac{2m_2}{m},~~~~
    \lambda=\frac{2L}{m},
    \label{ee6}
    \ee
where $m_1:=(m_+'+m_-')/2$, $m_2:=(m_+'-m_-')/2$, and $\lambda:=2\pi/k=2\pi L/\fK$ is the wavelength.

In summary, our device is invisible from both directions provided that the refractive indices of both the layers are rational numbers, and the wave length $\lambda$ at which the device is invisible is such that $L$ is a half integer multiple of $\lambda$, \footnote{The condition on $\lambda$ follows from the requirement that $\fn_1\neq\fn_2$. For $\fn_1=\fn_2$, invisibility from one or both directions implies that the slab is empty, $\fn_1=\fn_2=1$, in which case there is no restriction on $\lambda$.}. Notice that even bidirectional invisibility does not require $\cP\cT$-symmetry.

\section{Unidirectional Invisibility and $\cP\cT$-Symmetry}

Equations (\ref{e3a}) and (\ref{e4a}) are complex equations involving two complex variables, $\fn_\pm$, and a real variable $\fK$. Therefore, in principle, we can fix say the real part of $\fn_+$, which is bounded from below by $2$, and try to solve these equations for the imaginary part of $\fn_+$, the real and imaginary parts of $\fn_-$, and $\fK$. For the physically relevant range of the values of these variables this turns out to be difficult to implement directly. This calls for a more systematic study of the structure of (\ref{e3a}) and (\ref{e4a}). An important clue is provided by $\cP\cT$-symmetry of the invisibility equations that we established for general scattering potentials in Section~II. Here we offer a direct and explicit verification for the presence of this symmetry.

First we note that for our system the parity transformation means swapping the labels of $\fn_j$, i.e., $\fn_1\stackrel{\cP}{\longleftrightarrow} \fn_2$, and the time-reversal transformation corresponds to complex conjugation of the refractive indices, $\fn_j\stackrel{\cT}{\longrightarrow}\fn_j^*$. Therefore, $\fn_1\stackrel{\cP\cT}{\longleftrightarrow}\fn_2^*$, and consequently
    \be
    \fn_+\stackrel{\cP\cT}{\longrightarrow}\fn_+^*,~~~~~
    \fn_-\stackrel{\cP\cT}{\longrightarrow}-\fn_-^*,~~~~~
    \fa_+\stackrel{\cP\cT}{\longrightarrow}\fa_+^*,~~~~~
    \fa_-\stackrel{\cP\cT}{\longrightarrow}-\fa_-^*,~~~~~
    \tilde\fn_\pm\stackrel{\cP\cT}{\longrightarrow}\tilde\fn_\pm^*.
    \label{PT-}
    \ee
Performing these transformation on Equations~(\ref{e3a}) and (\ref{e4a}) and noting that their left-hand sides are analytic functions of $\fn_\pm,\fa_\pm$, and $\tilde\fn_\pm$, we see that $\cP\cT$-transformation maps these equations to their complex-conjugate. Therefore, the real and imaginary parts of the transformed equations are identical with those of (\ref{e3a}) and (\ref{e4a}). This confirms our general result ensuing the $\cP\cT$-symmetry of the invisibility equations. The presence of this symmetry suggests that we first examine the $\cP\cT$-symmetric solutions of (\ref{e3a}) and (\ref{e4a}).

\subsection{$\cP\cT$-Symmetric Invisible Configurations}

The optical system described by (\ref{n=}) is $\cP\cT$-symmetric provided that $\fn_1^*=\fn_2$. In this case, $\fn_+$ and $\fn_-$ are respectively real and imaginary. The same is true about $\fa_\pm$. In particular, if we denote by $\eta $ and $\kappa $ the real and imaginary parts of $\fn_1$, we have
    \be
    \fn_+=2\eta ,~~~~~\fn_-=2i\kappa ,~~~~~\fa_+=\fK\eta ,~~~~~
    \fa_-=i\fK\kappa ,~~~~~\tilde\fn_\pm=\eta^2+\kappa^2\pm 1=|\fn_1|^2\pm1.
    \label{pt1}
    \ee
Using these relations and the identities $\cos(i z)=\cosh z$ and $\sin(i z)=i\sinh z$, we can express (\ref{e3a}) and (\ref{e4a}) in the form
    \bea
    &&\left(\frac{\eta^2}{\eta^2+\kappa^2}\right)\cos(\fK\eta)+
    \left(\frac{\kappa^2}{\eta^2+\kappa^2}\right)\cosh(\fK\kappa)=\cos\fK,
    \label{e21}\\
    &&\frac{1}{2}
    \left[\left(1+\frac{1}{\eta^2+\kappa^2}\right)\eta\:\sin(\fK\eta)
    -\left(1-\frac{1}{\eta^2+\kappa^2}\right)\kappa\:\sinh(\fK\kappa)\right]=\sin\fK.
    \label{e22}
    \eea
It is easy to see that (\ref{e13}) holds automatically. Notice that (\ref{e21}) and (\ref{e22}) are real equations involving three real variables, $\eta$, $\kappa$, and $\fK$. This is an enormous simplification over the general non-$\cP\cT$-symmetric case that amounts to solving four real equations for five real unknowns.

Next, we recall that for typical optically active material, $|\kappa| \lesssim 10^{-3}$ and $1\leq\eta\lesssim 5$. Therefore $\kappa^2/\eta^2 \lesssim 10^{-6}$, and we can safely approximate $\eta^2+\kappa^2$ by $\eta^2$. This simplifies (\ref{e21}) and (\ref{e22}), and we find
    \bea
    &&\cos(\fK\eta)+
    \frac{\kappa^2\cosh(\fK\kappa)}{\eta^2}\approx \cos\fK,
    \label{e21a}\\
    &&\frac{1}{2}
    \Big[\left(1+\eta^{-2}\right)\eta\:\sin(\fK\eta)
    -\left(1-\eta^{-2}\right)\kappa\:\sinh(\fK\kappa)\Big]\approx\sin\fK.
    \label{e22a}
    \eea
Here and in what follows we use the symbol ``$\approx$'' to mean that we ignore terms of order $\kappa^2/\eta^2$.

Because $\fK=Lk=2\pi L/\lambda$, where $\lambda$ is the wavelength, $\fK$ can take very large values. However, according to (\ref{e21}), $\cosh(\fK\kappa)\leq 2\eta^2/\kappa^2$. This in turn implies
    \be
    \fK\leq \frac{2}{|\kappa|}\,\ln\left(\frac{2\eta}{|\kappa|}\right).
    \label{e24}
    \ee
Therefore the smaller $|\kappa|$ (or the gain/loss coefficient) is, the larger the upper bound on $\fK$ becomes. In particular, (\ref{e24}) does not impose any severe restriction on the wavelength at which the system is invisible from the left. For example, using the fact that usually $1\leq\eta\lesssim 5$ and $|\kappa| \lesssim 10^{-3}$, we find from (\ref{e24}) that
    \be
    \fK\lesssim 18420.
    \label{condi}
    \ee
This is equivalent to $L\lesssim 2930\lambda$. In particular, for $\lambda\geq 100~{\rm nm}$, we have $L \lesssim 293~\mu{\rm m}$, which can be easily realized experimentally.

Another simple consequence of (\ref{e21a}) is that $\eta=1$ implies $\kappa=0$. Again this corresponds to an empty slab. Therefore unlike for the $\cP\cT$-symmetric periodic potential considered in \cite{unidir}, the unidirectional invisibility can be achieved in our system provided that we use an optically active medium with $\eta>1$.

Next, we obtain an analytic solution of  (\ref{e21a}) and (\ref{e22a}). To do this we introduce
    \be
    X:=(\fK\kappa)^2,~~~~~~~\alpha:=\frac{\eta^2+1}{\eta^2-1},~~~~~~~
    \beta:=\frac{2\eta}{\eta^2-1},
    \ee
respectively solve  (\ref{e21a}) and (\ref{e22a}) for $\cosh(\fK\kappa)$ and $\sinh(\fK\kappa)$, and employ the identity $\cosh^2(\fK\kappa)-\sinh^2(\fK\kappa)=1$. This gives a quadratic equation in $X$ with a single positive solution. We can use this solution to express $\kappa$ in terms of $\eta$ and $\fK$ as follows.
    \be
    \kappa\approx\pm\,\eta\;\sqrt{
    \sqrt{\frac{1}{4}\big[\alpha\sin(\fK\eta)-\beta\sin\fK\big]^4+
    \big[\cos(\fK\eta)-\cos\fK\big]^2}-
    \frac{1}{2}\big[\alpha\sin(\fK\eta)-\beta\sin\fK\big]^2 }.
    \label{e26}
    \ee
Using this relation in (\ref{e21a}) we find an equation for $\eta$ and $\fK$. Because (\ref{e21a}) involves $\cosh$ with possibly large arguments, it is not suitable for numerical calculations. Therefore, we solve this equation for $\cosh(\fK\kappa)$ and employ the identity $\cosh^{-1}(z)=\ln(z\pm\sqrt{z^2-1})$ to express it as
    \be
    \fK|\kappa|-\ln\left\{\frac{\eta^2}{\kappa^2}\left(\cos\fK-\cos(\fK\eta)\pm
    \sqrt{\big[\cos\fK-\cos(\fK\eta)\big]^2-\frac{\kappa^4}{\eta^4}}\:\right)\right\}
    \approx  0.
    \label{e27}
    \ee
Now, we substitute (\ref{e26}) in the left-hand side of (\ref{e27}). The result is a function of $\eta$ and $\fK$ that we denote by $F_\pm(\eta,\fK)$. In this way, we can express (\ref{e27}) in the form
    \be
    F_\pm(\eta,\fK)\approx 0.
    \label{e28}
    \ee
We can choose some typical values for $\eta$ and solve (\ref{e28}) numerically to find the possible values of $\fK$. Once these are determined we use (\ref{e26}) to obtain the corresponding values of $\kappa$. In order for the latter to be experimentally meaningful, they should be sufficiently small. We expect this to be the case provided that the $\fK$ values we find fulfil the condition (\ref{condi}).

To be specific, suppose that the left-hand (right-hand) layer of the slab is the one containing the gain (lossy) material, so that $\kappa<0$. In this case, we find that $F_-(\eta,\fK)\neq 0$, while for each $\eta>0$ the equation $F_+(\eta,\fK)=0$ gives a (countably) infinite set of values for $\fK$. Using these values in (\ref{e26}) we can determine the corresponding values of $\kappa$. The $\fK$ and $\kappa$ values we obtain in this way satisfy (\ref{e28}), but they may not satisfy  (\ref{e21a}) and (\ref{e22a}). This is because in deriving (\ref{e28}) we affected functions on our expressions that are not one-to-one. Therefore, although every solution of  (\ref{e21a}) and (\ref{e22a}) is a solution of (\ref{e26}) and (\ref{e28}), the opposite is generally not true. In our numerical calculation, we found solutions of (\ref{e26}) and (\ref{e28}) and checked if they satisfy the exact equations (\ref{e21}) and (\ref{e22}). In this way we determined the acceptable approximate solutions of the problem that are valid whenever we can neglect $\kappa^2/\eta^2$. We then obtained the exact numerical solutions of  (\ref{e21}) and (\ref{e22}) in the neighborhood of the approximate solutions. For example for $\eta=1.4, 2.0, 3.4$ with $\fK$ ranging over $(1995,2005)$ we respectively found two, three, and five different values of $\lambda/L$ (between 317.7 and 318.9) for which the system displayed unidirectional invisibility. Except for a single data point, the approximate and exact results agreed to eight significant figures. We also checked the values of $M_{12}$ and $M_{22}$ to determine the directional of invisibility. It turns out that for the case that $\kappa<0$ (resp.\ $\kappa>0$) we have invisibility from the left (resp.\ right).

Note also that because the equations we used to characterize invisibility of our system are even in $\kappa$, changing the sign of the values of $\kappa$ that lead for invisibility from the left, we find the values that support invisibility from the right. This confirms the statement ii of the Invisibility Theorem.

\begin{figure}[h!]
    \begin{center}
    \includegraphics[scale=.7,clip]{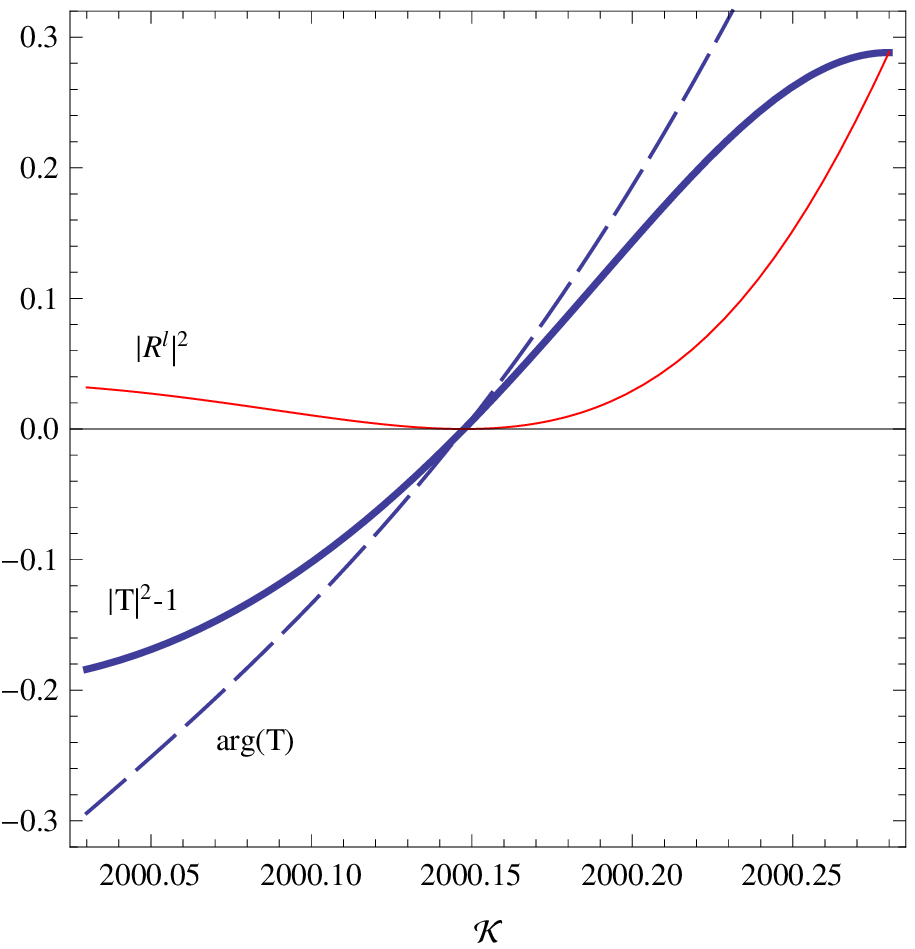}\vspace{-.1cm}
     \caption{(Color online) Graphs of $|T|^2-1$ (thick solid curve), argument (phase angle) of $T$ (dashed curve), and $|R^l|^2$ (thin solid red curve) as functions of $\fK=Lk=2\pi L/\lambda$ for a $\cP\cT$-symmetric two-layer slab with $\fn_1=\fn_2^*=3.4-0.003422i$. The intersection points of these curves and the $\fK$-axis corresponds to invisibility from the left for $\fK=2000.148$. For a sample consisting of two $150\:\mu{\rm m}$ thick layers, this gives the value $942.408$~nm for the wavelength.\label{fig1}}
    \end{center}
    \end{figure}

Figure~\ref{fig1} shows plots of $|T|-1$, $|R^l|$, and the argument (phase angle) of $T$ as a function of $\fK$ for a $\cP\cT$-symmetric sample with $\eta=3.4$ and $\kappa=-.00342163$. The fact that the graphs have a common intersection point located on the $\fK$-axis is a clear demonstration of the invisibility of the device from the left for the value of $\fK$ at this point, i.e., $\fK=2000.147552$. For these values of $\eta,\kappa$, and $\fK$, we have
    \be
    |R^l|^2<10^{-10},~~~~~~|T|^2-1<10^{-5},~~~~~~|R^r|^2>0.89.
    \label{boundzz}
    \ee
Taking the thickness of each layer of the slab to be 150~$\mu$m, so that $L=300\:\mu{\rm m}$, and using $\fK=2000.147552$, we find that the device displays invisibility from the left for $\lambda=942.408269$~nm.
If we keep only three decimal places in the above figures for $\kappa, \fK$ and $\lambda$, the first two bounds given in (\ref{boundzz}) change to $|R^l|^2<10^{-5}$ and $|T|^2-1<3\times 10^{-3}$, respectively.

\subsection{Non-$\cP\cT$-Symmetric Invisible Configurations}

In the absence of $\cP\cT$-symmetry, the equations governing unidirectional invisibility, (\ref{e3a}) and (\ref{e4a}), are complex equations involving two complex and one real unknowns, $\fn_+$, $\fn_-$, and $\fK$, respectively. It is more convenient to parameterize them in terms of $\fa_+$, $\fa_-$, and $\fK$.
This leads to
    \bea
    &&\fa_+^2(\cos\fa_+-\cos\fK)-\fa_-^2(\cos\fa_--\cos\fK)=0,
    \label{zee1}\\
    &&(\fa_+^2-\fa_-^2-\fK^2)\fa_+\sin\fa_++
    (\fa_+^2-\fa_-^2+\fK^2)\fa_-\sin\fa_--2(\fa_+^2-\fa_-^2)\fK\sin\fK=0.
    \label{zee2}
    \eea
Equating the real and imaginary parts of the left-hand side of these equations to zero, we find a set of extremely complicated real equations for $\fK$ and
    \be
    x_\pm:= \RE(\fa_\pm),~~~~~~~~
    y_\pm:= \IM(\fa_\pm).
    \label{e31}
    \ee
A direct Mathematica-based numerical treatment of these equations proves to be intractable for the physically relevant ranges of the values of $\fK$, $x_\pm$, and $y_\pm$. A careful study of the reasons for the failure of such a treatment guided us to a particular change of variables that allowed for a reliable perturbative method of solving these equations. The following is a brief outline of this method.
\begin{enumerate}
\item Because $|x_\pm|=|\RE(\fK\fn_\pm)|/2$ and $\fK$ are typically large numbers, we express them as
        \be
        x_\pm=2\pi m_\pm+\frac{\gamma_\pm}{2\pi m_\pm},~~~~~~~~~
        \fK=2\pi m_0+\frac{\gamma_0}{2\pi m_0},
        \label{new}
        \ee
    where $m_\pm$ and $m_0$ are integers, $m_+$ and $m_0$ are positive,
    and $\gamma_\pm$ and $\gamma_0$ are real parameters that are of the same order of magnitude as $y_\pm$. In particular, $|\gamma_\pm|,|\gamma_0|$ and $|y_\pm|$ are at least two orders of magnitude smaller than $m_+, |m_-|$, and $m_0$. In view of (\ref{new}) and the fact that
        \be
        \fa_\pm=x_\pm+i y_\pm,
        \label{new2}
        \ee
    this justifies the following approximations
        \bea
        &&\cos\fa_\pm\approx \cosh y_\pm-\frac{i\gamma_\pm\sinh y_\pm}{2\pi m_\pm},
        \label{zee3}\\
        &&\fa_\pm\sin\fa_\pm \approx \gamma_\pm\cosh y_\pm
        -y_\pm\sinh y_\pm+2\pi i m_\pm\sinh y_\pm,
        \label{zee4}\\
        &&\fa_\pm^2\approx 4\pi m_\pm(\pi m_\pm+i y_\pm),~~~~~~
        \cos\fK\approx 1,~~~~~\fK\sin\fK\approx \gamma_0.
        \label{zee5}
        \eea
    Note also that because $\fa_\pm=\fK\fn_\pm/2=\fK(\fn_1\pm\fn_-)/2$ and $\RE(\fn_1)\geq 1$ and $\RE(\fn_2)\geq 1$, Equations (\ref{new}) and (\ref{new2}) imply that
        \be
        |m_-| < m_+,~~~~~~~m_0\leq m_+.
        \label{condi-m}
        \ee

\item Substituting  (\ref{zee3}) -- (\ref{zee5}) in (\ref{zee1}) and (\ref{zee2}) and keeping only the dominant terms, we  respectively find from the real part of (\ref{zee1}) and the imaginary part of (\ref{zee2}) the following approximate equations.
    \bea
    \cosh y_-&\approx& \mu^2(\cosh y_+-1)+1,
    \label{zee11}\\
    \sinh y_-&\approx&-\mu\,\nu\sinh y_+,
    \label{zee12}
    \eea
where
    \be
    \mu:=\frac{m_+}{m_-},~~~~~\nu:=\frac{m_+^2- m_-^2+m_0^2}{m_+^2- m_-^2-m_0^2}.
    \label{mu-nu}
    \ee
Similarly implementing the approximations (\ref{zee3}) -- (\ref{zee5}) in the imaginary part of (\ref{zee1}) and the real part of (\ref{zee2}) and using (\ref{zee11}) and (\ref{zee12}) to simplify the resulting equations, we obtain
    \bea
    \mu\,\gamma_+\sinh y_+-\gamma_-\sinh y_-&\approx&
    2\mu (y_+-\mu y_-)(\cosh y_+-1),
    \label{zee13}\\
    \nu\,\gamma_+\cosh y_++\gamma_-\cosh y_-&\approx&
    [(\nu +\nu_+)y_+-\nu_0 y_-)\sinh y_+\nn\\
    &&+ [\nu_0 y_++(1 -\nu_-)y_-]\sinh y_-+(\nu_+-\nu_-)\gamma_0,
    \label{zee14}
    \eea
where
    \[\nu_\pm:=\frac{2m_\pm^2}{m_+^2- m_-^2-m_0^2},~~~~~~
    \nu_0:=\frac{2m_-m_+}{m_+^2- m_-^2-m_0^2}.\]

\item Inserting (\ref{zee11}) and (\ref{zee12}) in the identity $\cosh^2 y_- -\sinh^2y_-=1$ and using $\cosh^2 y_+-\sinh^2y_+=1$ to simplify the result, we find a quadratic equation in $\cosh y_+$ with solutions
        \bea
        \cosh y_+&\approx& 1,
        \label{zee15}\\
        \cosh y_+&\approx&\frac{\mu^2+\nu^2-2}{\mu^2-\nu^2}=
        \frac{1+\frac{\nu^2-1}{\mu^2-1}}{1-\frac{\nu^2-1}{\mu^2-1}}.
        \label{zee16}
        \eea
    In view of (\ref{zee11}), (\ref{zee15}) implies $y_\pm\approx 0$. This corresponds to the bidirectional invisibility that we examined in Section~IV. (\ref{zee16}) is acceptable provided that its right-hand side is greater than 1. This leads to the condition: $\nu^2<\mu^2$ which, in light of (\ref{condi-m}) and (\ref{mu-nu}), is equivalent to
        \be
        m_0<m_+-|m_-|.
        \label{condi-m2}
        \ee
   Supposing that this condition is fulfilled, we can invert (\ref{zee16}) to obtain
        \be
        y_+\approx \pm{\rm arccosh}\left(\frac{\mu^2+\nu^2-2}{\mu^2-\nu^2}\right).
        \label{zee17}
        \ee
   Furthermore, using (\ref{zee12}), (\ref{zee16}), and the identity: $\sinh y_+=\pm\sqrt{1-\cosh^2 y_+}$, we find \footnote{The functions ${\rm arccosh}$ and ${\rm arcsinh}$ appearing in (\ref{zee17}) and (\ref{zee18}) are respectively the principal part of $\cosh^{-1}$ and $\sinh^{-1}$ that take positive real values whenever they have a positive real argument and the argument of ${\rm arccosh}$ is greater than $1$.}
        \be
        y_-\approx \mp {\rm arcsinh}\left(\frac{2\mu\nu\sqrt{(\mu^2-1)(\nu^2-1)}}{\mu^2-\nu^2}\right).
        \label{zee18}
        \ee
   Finally, we substitute (\ref{zee17}) and (\ref{zee18})  in (\ref{zee13}) and (\ref{zee14}) to obtain a pair of linear equations for $\gamma_\pm$ and $\gamma_0$. We can easily solve any two of these variables in terms of the third.
\end{enumerate}

As we see our method includes four free parameters, namely $m_\pm$, $m_0$, and any one of $\gamma_\pm$ and $\gamma_0$. We can easily relate $m_\pm$ to the real part of $\fn_1$ and $\fn_2$, that we respectively denote by $\eta_1$ and $\eta_2$. More specifically,
    \be
    m_\pm\approx \frac{(\eta_1\pm\eta_2)m_0}{2}.
    \label{zee21}
    \ee
The integer $m_0$ is related to $\fK=Lk=2\pi L/\lambda$ according to
    \be
    m_0\approx\frac{\fK}{2\pi}=\frac{L}{\lambda}.
    \label{zee22}
    \ee
Therefore, for a given two-layer slab and any wavelength $\lambda$ we can
use the real part of the refractive index to determine $m_\pm$ and $m_0$. The choice of the last free parameter, i.e., one of $\gamma_\pm$ and $\gamma_0$ is arbitrary. We choose $\gamma_0$ to be a real  number of order 1 so that the corresponding $\gamma_\pm$ are also of order 1, \footnote{This actually does not restrict our choice of $\gamma_0$ significantly, because the dependence of $\gamma_\pm$ on $\gamma_0$ is linear with a slope and intercept of order 1.}. It turns out that the choice of the $+$ or $-$ sign in (\ref{zee17}) corresponds to invisibility from the right and left, respectively.

The following is the result of the application of our method for a sample with $\eta_1\approx 3.4$, $\eta_2\approx 1.4$, $\fK\approx 2000$, and $\gamma_0=-6$.
    \bea
    &&m_-=m_0=318,~~~~~~~~m_+=764,~~~~~~~~\fK=1998.049925, \\
    &&y_-=2.392197,~~~~~~~~~~y_+=-1.180878,
    \label{ys=}\\
    &&\gamma_-=-1.387129,~~~~~~~~\gamma_+=1.062178,\\
    &&\fn_1=3.402510 +i(6.062508\times 10^{-4}),~~~~
    \fn_2=1.402514-i(1.788281\times 10^{-3}).
    \label{non-pt=}
    \eea
For $L=300~\mu{\rm m}$, this sample displays invisibility from the left at
$\lambda=943.397644~{\rm nm}$. Inserting the above numerical values in the exact expressions for the reflection and transmission coefficients gives
    \[\big||T|^2-1\big|<2.1\times 10^{-5},~~~~~|{\rm arg}(T)|<3.2\times 10^{-3},
    ~~~~~|R^l|^2<2.8\times 10^{-6},~~~~~~|R^r|^2> 14.1.\]
Therefore our device is invisible from the left. The large value of $|R^r|^2$ indicates that it reflects the incident waves of this wavelength from the right after a 14-fold amplification. Furthermore, choosing the opposite sign for $y_\pm$ as in (\ref{ys=}) gives the time-reversed configuration:
    \bea
    &&\fn_1=3.402510 -i(6.062508\times 10^{-4}),~~~~
    \fn_2=1.402514+i(1.788281\times 10^{-3}),
    \eea
that is invisible from the right and has a large reflection coefficient from the left (that is also greater than 14.). This is a manifestation of the statement ii of the Invisibility Theorem.

\begin{figure}[t]
    \begin{center}
    \includegraphics[scale=1.1,clip]{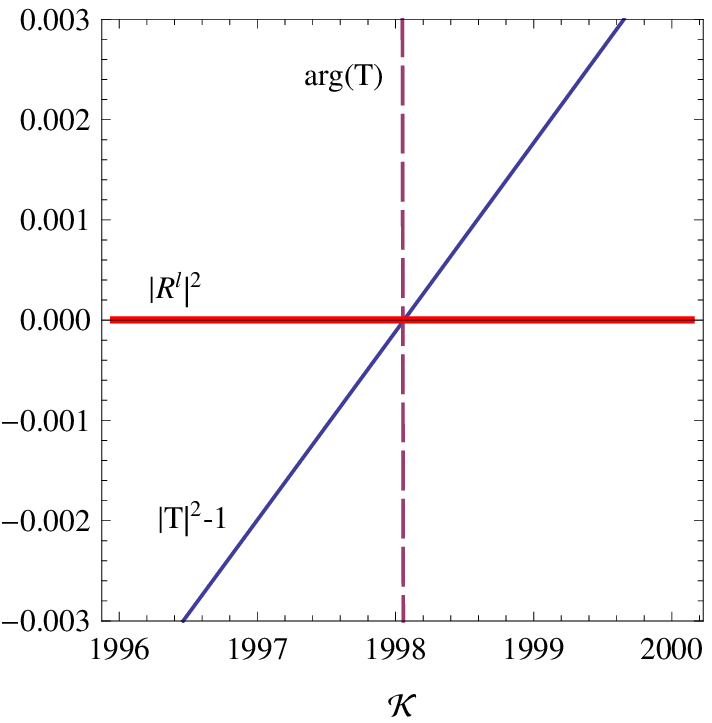} \hspace{.5cm}
    \includegraphics[scale=.77,clip]{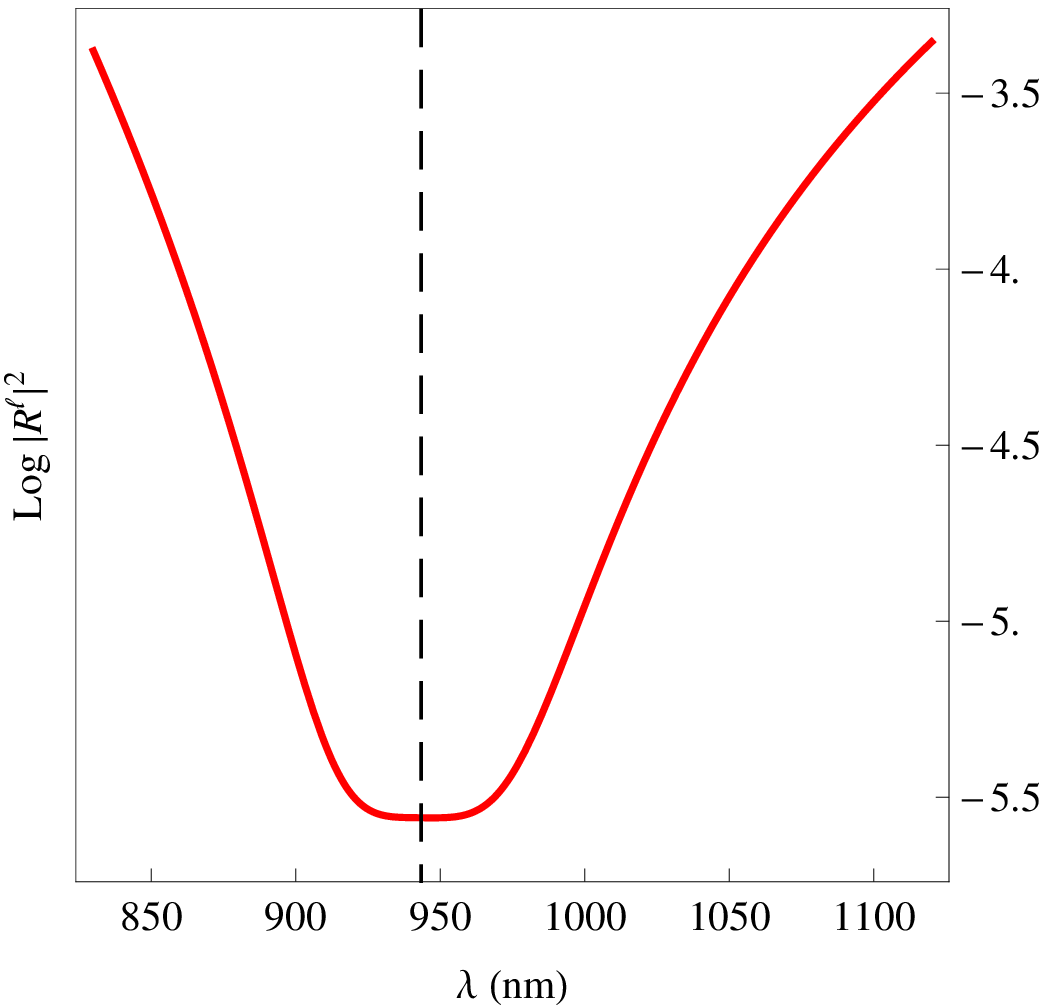}
    \vspace{-.1cm}
     \caption{(Color online) On the left are graphs of $|T|^2-1$ (solid blue curve), ${\rm arg}(T)$ (dashed purple curve), and $|R^l|^2$ (thick solid red curve) as functions of $\fK$ for a non-$\cP\cT$-symmetric two-layer slab with $\fn_1=3.403 +i(6.063\times 10^{-4})$ and $\fn_2=1.403-i(1.788\times 10^{-3})$. The intersection points of these curves and the $\fK$-axis corresponds to invisibility from the left for $\fK=1998.050$. For $L=300\:\mu{\rm m}$, this gives $\lambda=\lambda_\star:=943.398$~nm. On the right is the graph of $\log|R^l|^2$ as a function of the wavelength for $L=300\:\mu{\rm m}$. The dashed line represents $\lambda=\lambda_\star$ at which the system is invisible from the left. \label{fig2}}
    \end{center}
    \end{figure}

Figure~\ref{fig2} shows the plots of the $|R^l|^2$, $|T|^2-1$, and ${\rm arg}(T)$ as functions of $\fK$ and $\log|R^l|^2$ as a function of $\lambda$ for the non-$\cP\cT$-symmetric sample with specifications~(\ref{non-pt=}). An interesting outcome of our findings is that the reflection coefficient from the left remains negligibly small for a very wide spectral range. Although the invisibility from the left is lost for wavelengths slightly different from $\lambda=943.397644~{\rm nm}$, the system remains essentially reflectionless from the left (with $|R^l|^2<10^{-4}$) for wavelengths ranging from about $860~{\rm nm}$ to $1060~{\rm nm}$.

Next, we give the results for a sample with a different value of $\eta_2$. We take  $\eta_1\approx 3.4$, $\eta_2\approx 2.0$, $\fK\approx 2000$, and $\gamma_0=-2$. These lead to
    \bea
    &&m_0=318,~~~~~m_-=223,~~~~~m_+=859,~~~~~\fK=1998.051927,\nn \\
    &&y_-=1.704272,~~~~~y_+=-0.492952,~~~~~
    \gamma_-=0.612879,~~~~~\gamma_+=-1.307241,\nn\\
    &&\fn_1=3.402514 +i(6.062502\times 10^{-4}),~~~~
    \fn_2=1.999999-i(1.099683\times 10^{-3}),
    \label{n-n=3}\\
    &&\big||T|^2-1\big|<2.4\times 10^{-6},~~~~~|{\rm arg}(T)|<4.0\times 10^{-4},~~~~~|R^l|^2<7.0\times 10^{-7},~~~~~|R^r|^2> 0.9.~~~~~\nn
    \eea
For $L=300~\mu{\rm m}$ this configuration is invisible from the left at $\lambda=943.396699$~nm. Again changing the sign of the imaginary parts of $\fn_1$ and $\fn_2$, we find the dual configuration that is invisible from the right. Furthermore, for the original system (\ref{n-n=3}), the reflection coefficient from the left remains negligibly small (less than $10^{-4}$) for wavelengths between $780~{\rm nm}$ and $1165~{\rm nm}$. Therefore, the device is essentially reflectionless in a spectral range that is $385~{\rm nm}$ wide.

Finally, we should point out that if we keep only three decimal places in the numerical values we use in our calculations, the upper bounds we obtain for $\big||T|^2-1\big|$, $|{\rm arg}(T)|$, and $|R^l|^2$ are increased by about two orders of magnitude.

\section{Concluding Remarks}

The search for optical realizations of $\cP\cT$-symmetric potentials have unraveled unusual and interesting physical properties \cite{prl-2009,pt-optics}. Among these is the unidirectional invisibility achieved for the $\cP\cT$-symmetric locally periodic potential given by (\ref{lin=}), \cite{unidir}. In this paper, we showed that the phenomenon of the invisibility of a complex scattering potential is $\cP\cT$-symmetric in nature. By this we mean that irrespective of whether the potential possesses $\cP\cT$-symmetry or not the equations characterizing invisible potentials are $\cP\cT$-invariant. This does not means that non-$\cP\cT$-symmetric configurations cannot display unidirectional invisibility. It only means that the $\cP\cT$-symmetric invisible configurations are quite special, for they possess the same symmetry as the equations. A simple consequence of the presence of this symmetry is that the non-$\cP\cT$-symmetric configurations come in pairs that are related by the $\cP\cT$-transformation, i.e., they are $\cP\cT$-dual of one another. Furthermore, if a one-dimensional scattering system is invisible from the left for some value of the wavenumber $k$, then its time-reversal will be invisible from the right for the same value of $k$. These $\cP\cT$- and $\cT$-dual pairings of invisible potentials also apply to reflectionless potentials.

As a concrete implementation of our general results we have examined a simple two-layer planar slab of optically active material. We have studied in great detail the structure of the $\cP\cT$-symmetric as well as non-$\cP\cT$-symmetric invisible configurations of this model using essentially analytic techniques. In particular, we have shown by explicit calculations how $\cP\cT$-symmetry simplifies the characterization of the invisibility of our model. We have also confirmed the above-noted $\cT$-dual paring of invisible configurations by showing that the time-reversal of various left-invisible configurations are right-invisible. Furthermore, we found invisible configurations that remain reflectionless within a very wide spectral range. We intend to explore this phenomenon as a separate research project.
\vspace{3mm}

\noindent \textbf{{Acknowledgments:}} We wish to thank Aref Mostafazadeh and Ali Serpeng\"{u}zel for useful discussions. This work has been supported by the Scientific and Technological Research Council of Turkey (T\"UB\.{I}TAK) in the framework of the project no: 110T611, and by the Turkish Academy of Sciences (T\"UBA).

\ed

 \bea
    &&(x_-^2-y_-^2)\cos x_-\cosh y_-
    -(x_+^2-y_+^2) \cos x_+ \cosh y_+ \nn\\
    &&+2 x_- y_-  \sin x_-  \sinh y_-
    -2 x_+y_+  \sin x_+ \sinh y_+ =
    (x_-^2-x_+^2-y_-^2+y_+^2)\cos\fK,
    \label{e41}\\ &&\nn\\
    &&
    2 x_-y_-\cos x_-\cosh y_-
    -2 x_+y_+  \cos x_+ \cosh y_+
    \nn\\
    &&-(x_-^2-y_-^2) \sin x_- \sinh y_-
    +(x_+^2-y_+^2)\sin x_+\sinh y_+ =
    2 (x_-y_- -x_+y_+ ) \cos\fK,
    \label{e42}
    \eea
    \bea
    &&
    \left[x_-(\fK^2+x_-^2-x_+^2-3 y_-^2+y_+^2)+2x_+y_-y_+\right]\sin x_-\cosh y_- \nn\\
    &&-\left[x_+(\fK^2-x_-^2+x_+^2+y_-^2-3 y_+^2)+2 x_- y_- y_+\right]\sin x_+\cosh y_+
    \label{e43}\\
    &&-\left[y_-(\fK^2+3x_-^2-x_+^2-y_-^2+y_+^2)-2x_-x_+y_+\right]\cos x_-\sinh y_-
    \nn\\
    &&+\left[y_+(\fK^2-x_-^2+3x_+^2+y_-^2-y_+^2)-2x_-x_+y_-\right]\cos x_+\sinh y_+
    =2 \fK (x_-^2-x_+^2-y_-^2+y_+^2) \sin\fK,
    \nn\\
    &&\nn\\
    &&
    \left[y_-(\fK^2+3 x_-^2-x_+^2-y_-^2+y_+^2)-2x_-x_+y_+\right]\sin x_-\cosh y_-
    \nn\\
    &&-\left[y_+(\fK^2-x_-^2+3x_+^2+y_-^2-y_+^2)-2x_- x_+y_-\right]\sin x_+\cosh y_+
    \label{e44}\\
    &&+\left[x_-(\fK^2+x_-^2-x_+^2-3y_-2+y_+^2)+2x_+y_-y_+\right]
    \cos x_-\sinh y_-
    \nn\\
    &&-\left[x_+(\fK^2-x_-^2+x_+^2+y_-^2-3 y_+^2)+2x_-y_-y_+\right]\cos x_+\sinh y_+=4\fK(x_-y_--x_+y_+)\sin\fK.
    \nn
    \eea